\documentclass[a4paper,reqno]{amsart}
\numberwithin{equation}{section}
\usepackage{graphicx}
\usepackage{lscape}
\begin{document}

\newcommand{\undertilde}[1]{\underset{\widetilde{}}{#1}}
\newcommand{\al}{\alpha}
\newcommand{\de}{\delta}
\newcommand{\De}{\Delta}
\newcommand{\be}{\beta}
\newcommand{\ga}{\gamma}
\newcommand{\Ga}{\Gamma}
\newcommand{\Be}{\Xi}
\newcommand{\Al}{\Lambda}

\newcommand{\ala}{\bar{\alpha}}
\newcommand{\dea}{\bar{\delta}}
\newcommand{\bea}{\bar{\beta}}
\newcommand{\gaa}{\bar{\gamma}}

\newcommand{\als}{\hat{\alpha}}
\newcommand{\des}{\hat{\delta}}
\newcommand{\bes}{\hat{\beta}}
\newcommand{\gas}{\hat{\gamma}}

\newcommand{\alas}{\hat{\bar{\alpha}}}
\newcommand{\deas}{\hat{\bar{\delta}}}
\newcommand{\beas}{\hat{\bar{\beta}}}
\newcommand{\gaas}{\hat{\bar{\gamma}}}

\newcommand{\as}{\hat{a}}
\newcommand{\ds}{\hat{d}}
\newcommand{\bs}{\hat{b}}
\newcommand{\cs}{\hat{c}}

\newcommand{\ass}{\hat{\hat{a}}}
\newcommand{\dss}{\hat{\hat{d}}}
\newcommand{\bss}{\hat{\hat{b}}}
\newcommand{\css}{\hat{\hat{c}}}

\newcommand{\aas}{\hat{\bar{a}}}
\newcommand{\das}{\hat{\bar{d}}}
\newcommand{\bas}{\hat{\bar{b}}}
\newcommand{\cas}{\hat{\bar{c}}}

\newcommand{\la}{\lambda}
\newcommand{\La}{\Lambda}
\newcommand{\laa}{\bar{\lambda}}
\newcommand{\lab}{\bar{\bar{\lambda}}}
\newcommand{\lac}{\bar{\bar{\bar{\lambda}}}}
\newcommand{\lad}{\overset{4}{\lambda}}
\newcommand{\lae}{\overset{5}{\lambda}}
\newcommand{\laf}{\overset{6}{\lambda}}

\newcommand{\va}{\bar{v}}
\newcommand{\vs}{\hat{v}}
\newcommand{\vas}{\hat{\bar{v}}}

\newcommand{\vo}{v_1}
\newcommand{\voa}{\bar{v}_1}
\newcommand{\vos}{\hat{v}_1}
\newcommand{\voas}{\hat{\bar{v}}_1}

\newcommand{\vt}{v_2}
\newcommand{\vta}{\bar{v}_2}
\newcommand{\vts}{\hat{v}_2}
\newcommand{\vtas}{\hat{\bar{v}}_2}

\newcommand{\vth}{v_3}
\newcommand{\vtha}{\bar{v}_3}
\newcommand{\vths}{\hat{v}_3}
\newcommand{\vthas}{\hat{\bar{v}}_3}

\newcommand{\vf}{v_4}
\newcommand{\vfa}{\bar{v}_4}
\newcommand{\vfs}{\hat{v}_4}
\newcommand{\vfas}{\hat{\bar{v}}_4}

\newcommand{\xa}{\bar{x}}
\newcommand{\xb}{\bar{\bar{x}}}
\newcommand{\xc}{\bar{\bar{\bar{x}}}}
\newcommand{\xd}{\overset{4}{x}}
\newcommand{\xe}{\overset{5}{x}}
\newcommand{\xf}{\overset{6}{x}}

\newcommand{\xs}{\hat{x}}
\newcommand{\xss}{\hat{\hat{x}}}
\newcommand{\xas}{\hat{\bar{x}}}
\newcommand{\xbs}{\hat{\bar{\bar{x}}}}

\newcommand{\ya}{\bar{y}}
\newcommand{\yb}{\bar{\bar{y}}}
\newcommand{\ys}{\hat{y}}
\newcommand{\yss}{\hat{\hat{y}}}
\newcommand{\yas}{\hat{\bar{y}}}

\newcommand{\du}[3]{#1_{#2}^{#3}}
\newcommand{\ol}[1]{\overline{#1}}
\newcommand{\ul}[1]{\underline{#1}}
\newcommand{\os}[2]{\overset{#1}{#2}}

\newcommand\qon{{\rm qP}_{\rm{\scriptstyle I}}}
\newcommand\qtw{{\rm qP}_{\rm{\scriptstyle II}}}
\newcommand\qth{{\rm qP}_{\rm{\scriptstyle III}}}
\newcommand\qfi{{\rm qP}_{\rm{\scriptstyle V}}}
\newcommand\qsi{{\rm qP}_{\rm{\scriptstyle VI}}}

\newcommand\don{{\rm dP}_{\rm{\scriptstyle I}}}
\newcommand\dtw{{\rm dP}_{\rm{\scriptstyle II}}}
\newcommand\dth{{\rm dP}_{\rm{\scriptstyle III}}}

\newcommand\con{{\rm cP}_{\rm{\scriptstyle I}}}
\newcommand\ctw{{\rm cP}_{\rm{\scriptstyle II}}}

\newcommand{\eqn}[1]{(\ref{#1})}
\newcommand{\ig}[1]{\includegraphics{#1}}

\theoremstyle{break}    \newtheorem{Cor}{Corollary}
\theoremstyle{plain}    \newtheorem{Exa}{Example}[section]
\newtheorem{Rem}{Remark} \theoremstyle{marginbreak}
\newtheorem{Lem}[Cor]{Lemma}
\newtheorem{Def}[Cor]{Definition}
\newtheorem{prop}{Proposition}
\newtheorem{theorem}{Theorem}
\newtheorem{eg}{Example}
\newtheorem{fact}{Fact}

\title[2$\times$2 Lax pairs]{A completeness study on a class of discrete, 2$\times$2 {Lax pairs}}
\author{Mike Hay}
\address{School of Mathematics and Statistics F07, The University of Sydney, NSW 2006, Australia}
\email{mhay@mail.usyd.edu.au}

\begin{abstract}
We propose a method by which to examine all possible partial difference Lax pairs that consist of two 2$\times$2 discrete linear problems, where the matrices contain one separable term in each entry. We thereby derive new, higher-order versions of the lattice sine-Gordon and lattice modified KdV equations, while showing that there can be no other partial difference equations associated with this type of Lax pair.
\end{abstract}
\maketitle

\section{Introduction}
Nonlinear integrable lattice equations provide a natural discrete extension of classically integrable systems. An extensively studied example is the lattice modified Korteweg-de Vries equation, referred to as LMKdV:
\[
{\rm LMKdV:}\quad x_{l+1,m+1}=x_{l,m}\,
\frac{\bigl(p\,x_{l+1,m}-r\,x_{l,m+1}\bigr)}{\bigl(p\,x_{l,m+1}-r\,x_{l+1,m}\bigr)}
\]
where $x_{l,m}$ is the dependent variable, $l$ and $m$ are the discrete independent variables, and $p$ and $r$ are parameters. This equation provides an integrable discrete version of the well-known modified Korteweg-de Vries equation. Note that, in the past, the parameters $p$ and $r$ have been taken to be constant \cite{nc}, although singularity confinement has provided a way to de-autonomize LMKdV while maintaining its integrability, by allowing $p$ and $r$ to depend on $l$ and $m$ in a specific way \cite{pgr93}. The integrability of such equations lies in the fact that they can be solved through an associated linear problem called a Lax pair.

Lax pairs can appear in many guises, the type that we are exclusively concerned with in this paper consist of a pair of linear problems written:
\begin{equation}\label{linsys}
\begin{array}{ccc}
\theta(l+1,m)&=&L(l,m)\theta(l,m)\\
\theta(l,m+1)&=&M(l,m)\theta(l,m).
\end{array}
\end{equation}
where $\theta(l,m)$ is a two-component vector and $L(l,m)$ and $M(l,m)$ are $2\times2$ matrices. These linear problems are described by $L$ and $M$, which are referred to as the Lax matrices. If we were examining an explicitly determined Lax pair for LMKdV for example, the variable $x_{l,m}$, along with the parameters $p$ and $r$, would be found to reside within $L$ and $M$. The easily derived compatibility condition on this Lax pair is
\begin{equation}
L(l,m+1)M(l,m)=M(l+1,m)L(l,m)\nonumber
\end{equation}
and it is through this compatibility condition that we arrive at the integrable nonlinear equation associated with the Lax pair.

Despite the existence of a Lax pair often being used as the definition of integrability for a given equation \cite{bbt03}, there have been few studies that sought to find or categorize nonlinear equations that used Lax pairs as their starting point. Of those studies that did begin with Lax pairs, most chose a form of the Lax pair \emph{ a priori}, that is an assumption was made concerning the dependence of the linear systems on the spectral parameter, thus limiting the possible results.

The present study begins with Lax pairs that are 2$\times$2, where each entry of the Lax matrices contains only one separable term. The Lax pairs are otherwise general in that no assumptions are made as to the explicit dependence of any quantities within the Lax matrices on the lattice variables, $l$ and $m$, or on the spectral variable $n$. By one separable term we mean that each entry contains a term that can be split into a product of two parts, one that depends on the lattice variables, and another that depends only on the spectral variable. For example, in the $11$ entry of the $L$ matrix, we write the separable term $a(l,m)A(n)$. Both $a$ and $A$ may contain multiple terms themselves, say $a=\sum_i a_i$, $A=\sum_j A_j$, however, all terms within $a$ must multiply all those within $A$, the other entries their own similar term. So, we could have an $L$ matrix
\begin{equation}
L=\left(\begin{array}{cc}[a_1(l,m)+\ldots a_M(l,m)][A_1(n)+\ldots+A_N(n)]&b(l,m)B(n)\\c(l,m)C(n)&d(l,m)D(n)\end{array}\right)\nonumber
\end{equation}
where we have written the $11$ entry in the expanded form and left the other entries abbreviated to save space, no other terms can be added into any entry. $M$ must also contain just one separable term in each entry, although these terms are independent of those in $L$.

The reason for limiting the Lax pairs to those that are $2\times2$ with one separable term in each entry is two fold. Firstly, Lax pairs with more terms typically lead to equations of higher order, as can be seen from hierarchies of equations with Lax pairs \cite{hay2,jc}, therefore we constrain our study to the lower order equations by limiting the number of terms. Secondly, we limit the number of terms present in the compatibility condition and thus render it less complicated to examine all of the combinations of terms that can arise there. A combination of terms in the compatibility condition defines a system of equations that we subsequently solve, in a manner that preserves its full generality, up to a point where a nonlinear evolution equation is apparent, or it has been shown that the system cannot be associated with a nonlinear equation. Testing all combinations of terms, we thereby survey the complete set of Lax pairs of the type described.

In fact, of all the potential Lax pairs identified by this method, only two lead to interesting evolution equations. These are higher order varieties of the lattice sine-Gordon (LSG) and the LMKdV equations, which can be found in section \ref{results}. The remaining systems are shown to be trivial, overdetermined or underdetermined. As noted in \cite{srh}, this suggests a connection between the singularity confinement method and the existence of a Lax pair.

As we do not make any assumptions about the explicit dependence on the spectral parameter, we show that a particular nonlinear equation may have many Lax pairs, all depending on the spectral parameter in different ways. The effect that this freedom has on the process of inverse scattering is, as yet, unclear.

This paper is organized as follows: section \ref{results} presents the major results, those being the higher order versions of LMKdV and LSG, as well as a statement of the completeness theorem. The method of identifying and analyzing the viable Lax pairs is laid out in section \ref{links}, where a representative list of all the Lax pairs identified can be found. Section \ref{LMSG} explains how the higher order LSG and LMKdV equations are derived from the general form of their Lax pairs and section \ref{badevos} provides examples that describe why most Lax pairs found in section \ref{links} lead to trivial systems. A conclusion rounds out the paper.

\subsection{Results}\label{results}
Note that all difference equations in the remainder of this paper will utilize the notation
\begin{eqnarray}
\xa&=&x(l+1,m),\nonumber\\
\xs&=&x(l,m+1).\nonumber
\end{eqnarray}

As one of the two main results of the paper, we present two new integrable nonlinear partial difference equations. The first equation,
\begin{subequations}\label{resLSG}
\begin{eqnarray}
\text{{\rm LSG$_2$:}}\quad\quad\frac{\rho}{\sigma}\frac{\xs}{x}+\la_1 \mu_1\xas\ys&=&\frac{\sigma}{\rho}\frac{\xas}{\xa}+\frac{\la_2\mu_2}{x\ya} \\
\frac{\sigma}{\rho}\frac{\yas}{\ys}+\frac{\la_2\mu_2}{\xs y}&=& \frac{\rho}{\sigma}\frac{\ya}{y}+\la_1\mu_1\xa\yas
\end{eqnarray}
\end{subequations}
is referred to as LSG$_2$ because it is second order in each of the lattice dimensions, and because setting $x=y$ returns the familiar LSG equation, in a non-autonomous form. Here $x=x(l,m)$ and $y=y(l,m)$ are the dependent variables, $\la_i=\la_i(l)$ and $\mu_i=\mu_i(m)$ are parameter functions, as are $\rho=\la_3^{(-1)^m}$ and $\sigma=\mu_3^{(-1)^l}$.

Similarly, the equation
\begin{subequations}\label{resLMKdV}
\begin{eqnarray}
\text{{\rm LMKdV$_2$:}}\quad\quad\frac{\la_1}{\sigma}\frac{\xas}{\xs}+\frac{\mu_2}{\rho}\frac{y}{\ys}&=& \la_2\sigma\frac{y}{\ya}+\rho\mu_1\frac{\xas}{\xa}\\
\rho\mu_1\xs\yas+\la_2\sigma x\ys&=& \frac{\mu_2}{\rho}x\ya+\frac{\la_1}{\sigma}\xa\yas
\end{eqnarray}
\end{subequations}
where the terms are as for LSG$_2$, is referred to as LMKdV$_2$, again because setting $x=y$ brings about LMKdV. Note that the LMKdV so attained is of a more general form than the most common variety listed in the beginning of the introduction.

The second main result comes in the form of a theorem that we state as follows:

\begin{theorem} \label{theorem} The system of equations that arise via the compatibility condition of any $2\times2$ Lax pair \eqn{linsys} with one nonzero, separable term in each entry of each matrix (as described in the introduction) is either trivial, underdetermined, overdetermined, or can be reduced to one of LSG$_2$ or LMKdV$_2$.
\end{theorem}

The proof of theorem \ref{theorem} lies in considering all of the possible sets of equations that can arise from the compatibility condition of such Lax pairs, and solving those sets of equations in a way that retains their full freedom. This proof occupies the remainder of the paper.

The Lax pairs associated with LSG$_2$ and LMKdV$_2$ respectively are listed below, these Lax pairs are derived in section \ref{LMSG}. The Lax pair for LSG$_2$ is
\begin{subequations}\label{LSG2LP}
\begin{eqnarray}
L&=&\left(\begin{array}{cc}F_1/\rho&F_2\la_1\xa\\F_2\la_2/x&F_1\rho\ya/y\end{array}\right)\\
M&=&\left(\begin{array}{cc}F_2\xs/(\sigma x)&F_1/y\\F_1\mu_1\ys&F_2\sigma\end{array}\right)
\end{eqnarray}
\end{subequations}
While the Lax pair for LMKdV$_2$ is
\begin{subequations}\label{LMKdV2LP}
\begin{eqnarray}
L&=&\left(\begin{array}{cc}F_1\la_1\xa/x&F_2\rho/y\\F_2\ya/\rho&F_1\la_2\end{array}\right)\\
M&=&\left(\begin{array}{cc}F_1\mu_1\xs/x&F_2\ya/\sigma\\F_2\sigma\ya&F_1\mu_2\end{array}\right)
\end{eqnarray}
\end{subequations}
where $F_i=F_i(n)$ are arbitrary functions of the spectral variable $n$ with the condition that $F_1\neq kF_2$, where $k$ is a constant.

\section{Method of identifying potential Lax pairs} \label{links}
From Lax pairs of the type we consider here, the compatibility condition produces a set of equations, each equation being due to one of the linearly independent spectral terms that arises in some entry. Studies that have searched for integrable systems by beginning with a Lax pair, whether an isospectral or isomonodromy Lax pair or otherwise, typically assume some dependence of the Lax pair on the spectral parameter, then solve the compatibility condition for the evolution equation \cite{akns74,al76,as81,jbh92}. Most often a polynomial or rational dependence on the spectral variable is used \cite{w98}, but any type of explicit dependence could be investigated, for example Weierstrass elliptic functions.

\begin{eg} \label{eg}Consider the following $L$ and $M$ matrices
\begin{eqnarray}
L&=&\left(\begin{array}{cc}a\wp'&4b\wp\\
c&d\wp'\end{array}\right)\nonumber\\
M&=&\left(\begin{array}{cc}\al(\wp^{2}+1)&\be\wp'\\
\ga\wp'&\de\wp\end{array}\right)\nonumber
\end{eqnarray}
where $\wp$ is the Weierstrass elliptic function in the spectral parameter $n$ only, and all other quantities are functions of both the lattice variables $l$ and $m$. From the compatibility condition, $\widehat{L}M=\ol{M}L$, and noting that $\wp'^{2}=\frac{1}{4}\wp^{3}-g_{2}\wp-g_{3}$ where $g_{i}=constant$, we find the following equations in the $12$ entry
\begin{equation}
\begin{array}{crl}
\fbox{12}&\wp^{3}:&\,\,\as\be=b\ala+d\bea\\
&\wp^2:&\,\,\bs\de=0\\
&\wp:&\,\,\as\be=d\bea-\frac{b\ala}{4g_2}\\
&\wp^0:&\,\,\as\be=d\bea
\end{array}
\end{equation}
where we have separated out the equations coming from different orders of the spectral parameter.
\end{eg}

Clearly, this choice of $L$ and $M$ does not yield an interesting evolution equation through their compatibility, this example was instead chosen because it illustrates an important point. Multiplying together two functions of the spectral parameter can produce numerous orders that may or may not be proportional to other spectral term products in the compatibility condition. From example \ref{eg}, the spectral term multiplying $\wp^{2}$ did not `match up' with other terms in the $12$ entry at that order, which brought about the equation $\bs\de=0$, forcing some term to be zero. However, the other spectral terms did turn out more meaningful equations, highlighting the need to choose the dependence on the spectral parameter carefully so that none of the resulting equations force any lattice terms to be zero.

The inclusion of a zero lattice term does not preclude the existence of an interesting evolution equation. However, we are essentially classifying Lax pairs by the number of terms in their entries and the class of Lax pairs presently under inspection contains one separable term in each entry of their $2\times2$ matrices. If any of those terms were forced to zero then the resulting Lax pair would actually come under a different category in the present framework.

The most general form of a $2\times2$ Lax pair with exactly one separable term in each entry of the $L$ and $M$ matrices is:
\begin{eqnarray}
L&=&\left(\begin{array}{cc}aA&bB\\
cC&dD\end{array}\right)\nonumber\\
M&=&\left(\begin{array}{cc}\al\Lambda&\be\Xi\\
\ga\Gamma&\de\Delta\end{array}\right)\nonumber
\end{eqnarray}
Where lower cases represent lattice terms and upper cases represent spectral terms. The compatibility condition is $\widehat{L}M=\ol{M}L$, of which we initially concentrate on the $12$ entry.
\begin{equation}
\as \be A\Be+\bs\de B\De=b\ala B\Al+d\bea D\Be
\end{equation}

At this stage we are only concerned with the various linearly independent spectral terms that appear. These will determine the set of equations that come out of the compatibility condition, which are subsequently solved to find the corresponding evolution equation. That being the case, there are four quantities to contend with in this entry, $A\Be$, $B\De$, $B\Al$ and $D\Be$. Any of these four products can lead to multiple, linearly independent spectral terms, all of which must match up with at least one other spectral term from one of the three remaining products in this entry. If there exists some spectral term that does not match up with a spectral term from another product, then the lattice term that multiplies it will have to be zero, which is forbidden.

Let us label the terms in each product as follows $A\Be=\sum_i F_{A\Be_{i}}$, $B\De=\sum_i F_{B\De_{i}}$, etc. All the spectral terms that occur in the $12$ entry of the compatibility condition can be sorted into four groups according to the lattice terms that they multiply.
\begin{equation}
\begin{array}{ccccccccccc}
\fbox{12}& &(\as\be)&    &(\bs\de)     & &|  &(b\ala)            &(d\bea)& &\\
         & &F_{A\Be_{1}}&&F_{B\De_{1}} & &|  &F_{B\Al_{1}}       &F_{D\Be_{1}}& & \\
         & &F_{A\Be_{2}}&&F_{B\De_{2}} & &|  &F_{B\Al_{2}}       &F_{D\Be_{2}}      & & \\
         & &\vdots      &&\vdots       & &|  &\vdots             &\vdots&&
\end{array}\label{groups12}
\end{equation}
Where the line separating the four groups marks the position of the equals sign in the associated lattice term equations. Organizing the terms from example \ref{eg} in this way leads to
\begin{equation}
\begin{array}{ccccccccccc}
\fbox{12}& &(\as\be)&    &(\bs\de)     & &|  &(b\ala)     &(d\bea)& &\\
         & &\wp^3&       & \wp^2& &       |  &\wp^3       &\wp^3& & \\
         & &\wp&         &      & &       |  &\wp         &\wp      & & \\
         & &\wp^0      & &      & &       |  &            &\wp^0&&
\end{array}\nonumber
\end{equation}

The present study will utilize the word `group' in the general English sense, our groups refer to collections of spectral terms that multiply the same lattice terms in an entry of the compatibility condition. We have already seen that all terms in all groups must be proportional to a term from at least one of the other three groups in this entry. Conversely, where there are spectral terms that are proportional to others from other groups, an equation relating the corresponding lattice terms will thus be defined. Still with the above example, that Lax pair has a dependence on the spectral parameter such that the groups multiplying $\as\be$, $b\ala$ and $d\bea$ all contain the spectral term $\wp^3$, for which the corresponding equation in the lattice terms is $\as\be=b\ala+d\bea$.

At this stage, the number of possible sets of proportional spectral terms, and therefore the number of possible combinations of equations yielded by the compatibility condition, is unmanageably large. We require further considerations to bring the problem under control.

\subsection{Links and equivalent equations at different orders}
The following definition applies to sets of spectral terms within an entry of the compatibility condition that bring about the equations used to find the evolution equation. The word \emph{proportional}\, is used in the sense that two terms $F_1$ and $F_2$ are proportional if $F_1=kF_2$ for some finite constant $k$.

\begin{Def}
A \textbf{link} is a set of proportional spectral terms in the same matrix entry of the compatibility condition. A set of two proportional terms is a single link, three terms a double link, and four terms a triple link.
\end{Def}
Naturally, the spectral terms that comprise a link must each reside in a different group of spectral terms within an entry. If there are two or more spectral terms that are proportional to one another within the same group, they are simply added together to make one term. Since each group of spectral terms multiplies the same lattice term, one may speak of either links between the groups of spectral terms or links between lattice terms, with the same meaning. The above definition captures the idea that the entries of the compatibility condition give rise to different lattice term equations at different orders in the spectral term, without appealing powers of some basic function.

With the employment of proportional terms comes the possibility of constants of proportionality, which we begin to deal with here.

\begin{fact}\label{fact} If there exist two distinct single links between the same two groups of spectral terms, then the corresponding constants of proportionality must be equal. \end{fact}

The proof of this fact is elementary: say that one single link is formed by allowing $F_{A\Be_1}\propto F_{D\Be_1}$, where neither of these terms is proportional to any other spectral term that arises in the $12$ entry, and the other single link corresponds to $F_{A\Be_2}\propto F_{D\Be_2}$, where again these terms link with no others. By including some constants of proportionality, $k_1$ and $k_2$ respectively, we can write down the equations that correspond to these two single links, those being $\as\be=k_1 d \bea$ and $\as\be=k_2 d \bea$. Since both equations must hold, and none of the lattice terms can be zero, it is clear that we must have $k_1=k_2$.

This simple fact proves to be rather important because it ensures that all links between the same two groups can be bundled together. Further, all the spectral terms, in some group, that correspond to those single links with one other group, can be treated as a single spectral term. So, where $F_{A\Be_1}$ and $F_{D\Be_1}$ formed one single link and $F_{A\Be_2}$ and $F_{D\Be_2}$ formed another between the same two groups, we can lump together $F_{A\Be_1}+F_{A\Be_2}=G_1$ and be sure that it links with $F_{D\Be_1}+F_{D\Be_2}=kG_1$.

Still with the $12$ entry, if there exist multiple double links between the same three lattice terms, $\as\be$, $b\ala$ and $d\bea$ say, then the lattice term equations that result from those links can be written
\begin{equation}
K\left(\begin{array}{c}\as\be\\b\ala\\d\bea\end{array}\right)=0 \label{doubleconstants}
\end{equation}
Where $K$ is a matrix of the constants of proportionality between the various spectral terms, normalized so that each entry of the first column of $K$ is unity. Each row of $K$ corresponds to a double link. If $K$ is such that equation \eqn{doubleconstants} is over-determined or uniquely solvable, then any Lax pair possessing the corresponding links is inconsistent or contains a zero lattice term. Therefore, we need not consider more than two double links between the same three spectral terms, although there may be multiplicity within those two double links. By the same argument we can allow a maximum of three different triple links between the same four lattice terms in an entry of the compatibility condition.

\subsection{Link symbolism}
The abundance of Lax pairs that need to be checked necessitates the introduction of a shorthand, which will be based on their links. The off-diagonal entries both contain four groups of spectral terms, each of which multiplies a single product of lattice terms. The $12$ entry contains the spectral term products $A\Be$, $B\De$, $B\Al$, and $D\Be$ associated with the lattice term products $\as\be$, $\bs\de$, $b\ala$ and $d\bea$ respectively. For the shorthand, we always set out the spectral terms in the same way on the page
\begin{equation}
\begin{array}{cc}A\Be&B\Al\\B\De&D\Be\end{array}\nonumber
\end{equation}
Each link can be represented by lines between the quantities that are proportional to each other, and so we will use the symbols listed in table \ref{offdiagsymbols} to represent the combinations of links in the $12$ entry, where $F_{A\Be}$ is some term from the group of spectral terms formed by taking the product $A\Be$, and other terms are similarly labeled.
\begin{table}[!h]
\begin{center}
\begin{tabular}{|c|c|}
\hline Symbol& Links\\
\hline \includegraphics{cross.1}& $F_{A\Be}\propto F_{D\Be}$, $F_{B\De}\propto F_{B\Al}$\\
 \includegraphics{vert.1}& $F_{A\Be}\propto F_{B\De}$, $F_{B\Al}\propto F_{D\Be}$\\
 \includegraphics{horiz.1}& $F_{A\Be}\propto F_{B\Al}$, $F_{B\De}\propto F_{D\Be}$\\
 \includegraphics{all.1}&$F_{A\Be}\propto F_{D\Be}\propto F_{B\De}\propto F_{B\Al}$\\
 \includegraphics{overt.1}&$F_{A\Be}\propto F_{B\De}\propto F_{B\Al_i}$, $F_{B\Al_j}\propto F_{D\Be}$\\
 \vdots&\vdots\\\hline
\end{tabular}
\caption{Symbols used to represent the link combinations in the off-diagonal entries. Note that $F_{B\Al_i}\neq kF_{B\Al_j}$, $k$ a constant}\label{offdiagsymbols}
\end{center}
\end{table}

The $21$ entry is similar to the $12$ entry in that it possesses four distinct products of spectral terms. The same symbols listed in table \ref{offdiagsymbols} are used again for the $21$ entry, with clear meaning given that the spectral products are set out as follows
\begin{equation}
\begin{array}{cc}C\Al&A\Ga\\D\Ga&C\De\end{array}\nonumber
\end{equation}

However, the diagonal entries are slightly different because each diagonal entry contains a product that occurs twice: $A\Al$ occurs twice in the $11$ entry and $D\De$ twice in $22$. This automatically causes the associated lattice terms to be paired in their respective entries and, as such, $A\Al$ and $D\De$ need not be linked with another spectral term to prevent a zero lattice term. This being the case, there are really only three spectral term products to consider in both of the diagonal entries, one of which need not be linked to the other two, and our symbols reflect that. Positioning the spectral term products as follows
\begin{equation}
A\Al\quad\begin{array}{c}
B\Ga\\
C\Be\end{array}\nonumber
\end{equation}
We symbolize the links as indicated in table \ref{diagsymbols}, where the symbol `\ig{double.1}' is used to represent the repeated spectral term products.

\begin{center}
\begin{table}[h!]
\begin{tabular}{|c|c|}
\hline Symbol& Links\\
\hline \includegraphics{dall.1}& $F_{A\Al}\propto F_{B\Ga} \propto F_{C\Be}$\\
 \includegraphics{pairs.1}&  $F_{A\Al}$ alone, $F_{B\Ga} \propto F_{C\Be}$\\
 \includegraphics{split.1}& $F_{A\Al_i}\propto F_{B\Ga}$ and $F_{A\Al_j} \propto F_{C\Be}$ separately\\
 \vdots&\vdots\\\hline
\end{tabular}
\caption{Symbols used to represent the link combinations in the diagonal entries of the compatibility condition. Note that $F_{A\Al_i}\neq kF_{A\Al_j}$, $k$ a constant}\label{diagsymbols}
\end{table}
\end{center}

\subsection{Which Lax pairs need to be checked?}
All link combinations that do not force any lattice terms to be zero are checked systematically. The procedure for doing this runs as follows:
\begin{enumerate}
\item Begin with the $12$ entry, assume that only single links exist there, and list all single link combinations that produce different lattice term equations.
\item Construct the links found in the previous step by choosing proportional sets of spectral terms in the appropriate groups.
\item Move to the $21$ entry, noting the spectral term constructions from the previous step, and identify all the viable link combinations in this entry.
\item Repeat the previous step in the diagonal entries.
\end{enumerate}

After identifying all link combinations with single links in the $12$ entry, we repeat the entire process assuming that double and possible single links exist there, and then repeat once more with triple, and possibly double and single links in the $12$ entry. Finally, the corresponding set of lattice term equations for each Lax pair must be analyzed to find the resulting evolution equation. This analysis needs to be conducted in a manner that preserves the full freedom of the system, as described in section \ref{LMSG}.

\subsubsection{Single links in the $12$ entry}
The four groups of spectral terms in the $12$ entry are
\begin{equation}
\begin{array}{cc}A\Be&B\Al\\B\De&D\Be\end{array}\nonumber
\end{equation}
Each group must be linked to another so, using single links, the group $A\Be$ must be linked to at least one of the three other groups, $B\Al$, $B\De$, or $D\Be$. For arguments sake, say that there exists a single link between $A\Be$ and $B\Al$.
\begin{equation}
\begin{array}{ccc}
A\Be& &B\Al\\
&\ig{top.1}& \\
B\De& &D\Be
\end{array}\nonumber
\end{equation}
That leaves both groups $B\De$ and $D\Be$ requiring links and, since we are only concerned with single links at the moment, these two groups of spectral terms can be linked to each other, or to one of $A\Be$ or $B\Al$, in distinct, single links. However, it can be shown that if any group in either off-diagonal entry possesses single links between it and two other groups, the resulting Lax pair is associated with a trivial evolution equation (see proposition \ref{twosinglesoneentry} below). Hence, given the first link between $A\Be$ and $B\Al$, the only other single link that needs to be considered is between $B\De$ and $D\Be$. Therefore, table \ref{singlecombos} lists the only single link combinations in the $12$ entry that require further analysis.
\begin{table}[h!]
\begin{center}
\begin{tabular}{c c c c c c}
\fbox{12}&$\begin{array}{cc}A\Be&B\Al\\B\De&D\Be\end{array}$&\quad&\ig{horiz.1}&\ig{vert.1}&\ig{cross.1}
\end{tabular}\caption{Single link combinations in the $12$ entry}\label{singlecombos}
\end{center}
\end{table}

\begin{prop}\label{twosinglesoneentry}
If there exist two single links between some lattice term and two others in an off-diagonal entry, then the resulting evolution equation is trivial.
\end{prop}
The proof of Proposition \ref{twosinglesoneentry} lies simply in checking all the possible link combinations that meet the criterion. In that sense there is nothing special about Lax pairs of this type, they are only treated separately because there are many such cases, so including them with the others would unnecessarily lengthen the argument.

Before links in the $21$ entry are examined, we must select spectral terms that engender the links already chosen in the $12$ entry. We shall proceed with the analysis under the assumption that single links between $A\Be$ and $B\Al$, and between $B\De$ and $D\Be$, although the other combinations can be dealt with in the same way. The specified links are constructed by choosing spectral terms
\begin{subequations}\label{hhlrST}
\begin{eqnarray}
A&=&\frac{1}{\Be}(F_1+\ldots)\\
D&=&\frac{1}{\Be}(F_2+\ldots)\\
\Al&=&\frac{1}{B}(F_1+\ldots)\\
\De&=&\frac{1}{B}(F_2+\ldots)
\end{eqnarray}
\end{subequations}
where $F_i=F_i(n)$ and $F_1\neq kF_2$, $k$ a constant. It is understood that while there is room for other spectral terms in the expression $A=\frac{1}{\Be}(F_1+\ldots)$, there cannot be a term proportional to $F_2/\Be$ in $A$, as this would cause the single link corresponding to $F_2$ to become a double link. In this way the desired links are constructed, plus we have allowed for additional links should they be appropriate or required later. Readers may note the omission of any constants of proportionality in the above, however, the constants that could have been written at this point can all be absorbed into the lattice terms that they multiply.

Turning our attention to the $21$ entry of the compatibility condition, $\cs\al C\Al+ \ds\ga D\Ga=a\gaa A\Ga+c\dea C\De$, the following spectral terms appear
\begin{equation}
\begin{array}{ccccccccccc}
\fbox{21}& &(\cs\al)&    &(\ds\ga)     & &|  &(a\gaa)            &(c\dea)& &\\
         & &\frac{C}{B_{ _{}}}F_{1}&&\frac{\Ga}{\Be}F_{2} & &|  &\frac{\Ga}{\Be}F_{1}&\frac{C}{B}F_{2}& & \\
         & &\vdots&&\vdots & &|  &\vdots      &\vdots      & &
\end{array}\label{hhlrgroups21}
\end{equation}

Since every spectral term must be proportional to another in the same entry, $\frac{C}{B}F_{1}$ must be proportional to $\frac{\Ga}{\Be}F_{1}$ or $\frac{\Ga}{\Be}F_{2}$. Clearly, $\frac{C}{B}F_{1}$ cannot be proportional to $\frac{C}{B}F_{2}$, nor can it link with some other term that we are yet to define, as this would introduce single links between some lattice term and two others, the situation excluded by proposition \ref{twosinglesoneentry}. Moreover, we can exclude the case with $\frac{C}{B}F_{1}\propto\frac{\Ga}{\Be}F_{2}$ because this also requires the remaining spectral terms to be proportional to one another, \emph{i.e.} $\frac{C}{B}F_{2}\propto\frac{\Ga}{\Be}F_{1}$. These two conditions on the spectral terms imply that $F_1\propto F_2$, contradicting a previous assumption. Hence, the links chosen in the $12$ entry leave only one choice for the links in the $21$ entry of the compatibility condition: $\frac{C}{B}F_{1}\propto\frac{\Ga}{\Be}F_{1}$ and $\frac{C}{B}F_{2}\propto\frac{\Ga}{\Be}F_{2}$, which can be written more succinctly as $C\Be=B\Ga$
\begin{equation}\nonumber
\begin{array}{ccccccccccc}
\fbox{12}&A\Be&            &B\Al&\quad&&\quad&\fbox{21}&C\Al&            &A\Ga\\
         &    &\ig{horiz.1}&    &&\Rightarrow&&        &    &\ig{horiz.1}&    \\
         &B\De&            &D\Be&&           &&        &D\Ga&            &C\De
\end{array}
\end{equation}

The terms arising in the diagonal entries are listed below
\begin{equation}
\begin{array}{ccccccccc}
\fbox{11}&&&& & &| && \\
&&(\as\al-a\ala)&F_1^2/(B\Be)& & &| &(c\bea-\bs\ga)&B\Ga \\
&&&& & &| && \\
 &&&& & & && \\
\fbox{22}&&&& & &| && \\
&&(\ds\de-d\dea)&F_2^2/(B\Be)& & &| &(b\gaa-\cs\be)&B\Ga \\
&&&& & &| &&
\end{array}\label{hhlrgroups11}
\end{equation}
Note that the terms in equation \eqn{hhlrgroups11} are set out slightly differently to those in table \ref{diagsymbols} because we have already determined that $B\Ga\propto C\Be$.

The spectral terms in the diagonal entries, in this case, do not necessarily have to link with others because they are multiplied by more than one lattice term, as indicated in equation  \eqn{hhlrgroups11} where the lattice terms appear in parentheses to the left of the spectral terms they multiply. A lone spectral term in the diagonal entries, given the links already constructed in the off-diagonal entries, will not bring about a zero lattice term. Consequently, there exist three possible link combinations for the diagonal entries: $B\Ga\propto F_1^2/(B\Be)$, $B\Ga\propto F_2^2/(B\Be)$ or $B\Ga$ is proportional to neither $F_2^2/(B\Be)$ nor $F_1^2/(B\Be)$. All three choices lead to trivial evolution equations and we shall continue the analysis under the assumption that $B\Ga\propto F_1^2/(B\Be)$. A link combination has now been chosen in each of the entries of the compatibility condition, these are shown in table \ref{eglinks}
\begin{table}[h!]
\renewcommand{\arraystretch}{2}
\begin{center}
\begin{tabular}{|c| c |c |c|}
\hline
$\begin{array}{c} \fbox{12}\\
\begin{array}{cc}\mbox{\footnotesize{$A\Be$}}&\mbox{\footnotesize{$B\Al$}}\\
\mbox{\footnotesize{$B\De$}}&\mbox{\footnotesize{$D\Be$}}\end{array}\end{array}$ &$\begin{array}{c} \fbox{21}\\
\begin{array}{cc}\mbox{\footnotesize{$C\Al$}}&\mbox{\footnotesize{$D\Ga$}}\\
\mbox{\footnotesize{$A\Ga$}}&\mbox{\footnotesize{$C\De$}}\end{array}\end{array}$&$\begin{array}{c} \fbox{11}\\
\mbox{\footnotesize{$A\Al$}}\begin{array}{c}\mbox{\footnotesize{$B\Ga$}}\\
\mbox{\footnotesize{$C\Be$}}\end{array}\end{array}$ &$\begin{array}{c} \fbox{22}\\
\mbox{\footnotesize{$D\De$}}\begin{array}{c}\mbox{\footnotesize{$B\Ga$}}\\
\mbox{\footnotesize{$C\Be$}}\end{array}\end{array}$\\
\hline\ig{horiz.1}&\ig{horiz.1}&\ig{dall.1}&\ig{split.1}\\
\hline
\end{tabular}\caption{An example of the links that define a Lax pair}\label{eglinks}
\end{center}
\end{table}

Gauge transformations can be used to remove the dependence on some of the spectral terms and, as such, we expect some redundancy. The links constructed above are achieved by setting the values of the spectral terms to
\begin{equation}\nonumber
\begin{array}{ccccccc}
A&=&F_1,&\quad\quad&\Al&=&F_1\\
B&=&1,&\quad\quad&\Be&=&1\\
C&=&F_1^2,&\quad\quad&\Ga&=&F_1^2\\
D&=&F_2,&\quad\quad&\De&=&F_2\\
\end{array}
\end{equation}
where $F_1$ and $F_2$ are any functions of the spectral variable $n$, such that $F_1\neq kF_2$, $k=$ constant. Note that one of $F_1$ or $F_2$ may itself be a constant. No constants of proportionality are required as these can be absorbed into lattice terms in this case. The same links are reproduced by any suite of spectral terms that satisfies the following conditions, brought about by the links described above
\begin{equation}\nonumber
\begin{array}{ccccccc}
A&=&F_1/\Be,&\quad\quad&\Al&=&F_1/B\nonumber\\
C&=&F_1^2/\Be,&\quad\quad&\Ga&=&F_1^2/B\nonumber\\
D&=&F_2/\Be,&\quad\quad&\De&=&F_2/B\nonumber
\end{array}
\end{equation}

The resulting set of equations that are produced by the compatibility condition for this Lax pair are written in \eqn{hhlreqns}, although they can be read from the links given in table \ref{eglinks}.
\begin{equation}\label{hhlreqns}
\begin{array}{rcl}
\as\al-a\ala&=&c\bea-\bs\ga\\
\ds\de-d\dea&=&0\\
b\gaa-\cs\be&=&0\\
\as\be&=&b\ala\\
\bs\de&=&d\bea\\
\cs\al&=&a\gaa\\
\ds\ga&=&c\dea
\end{array}
\end{equation}

This rounds the description of the Lax pairs with only single links in the off-diagonal entries, in practice one would continue by analyzing equations \eqn{hhlreqns} to find the associated evolution equation, which in this case is trivial. On the possibility of including extra spectral terms to augment the links used here, see equation \eqn{hhlrST}, we note that fewer terms allow greater freedom and that any additional links could only lead to a more constrained system, one that certainly could not sustain an interesting evolution equation considering that the less constrained example here leads to a trivial result. Also, the alternative links that cause there to be one equation in the $22$ entry and two equations in the $11$ entry of the compatibility condition, see after equation \eqn{hhlrgroups11}, lead to the same evolution equations found here as the Lax pair is symmetric in that sense.

\subsubsection{Double links in the $12$ entry}
Here link combinations that consist of double and possibly single links in the $12$ entry are investigated. There are more possibilities in this class than when only considering single links, however the number is reduced by noting that some sets of link combinations are equivalent from the perspective of the lattice term equations. For example the following pair of link combinations in the $12$ entry are clearly equivalent.
\begin{equation}\nonumber
\begin{array}{cccccc}
\fbox{12}&\begin{array}{cc}A\Be&B\Al\\B\De&D\Be\end{array}&\quad&\ig{ohoriz.1}&\,\leftrightarrow\,&\ig{thoriz.1}\\
\end{array}
\end{equation}
Also, all link combinations that possess two double links in the $12$ entry are nearly equivalent, enough to consider them all together. The equivalence is because the lattice term equations corresponding to any two double links in this entry can be manipulated so that the they are the same as those from any other combination of two double links. The difference that may arise comes from the diagonal entries, where the particular pair of double links chosen in the $12$ entry can affect the variety of link combinations possible. However, the difference is not sufficient to alter the overall outcome that these systems are overdetermined. The complete list of double link combinations in the $12$ entry is listed in table \ref{alldoubles}
\begin{center}
\begin{table}[!h]
\begin{tabular}{c c c c c c c c c}
\fbox{12}&$\begin{array}{cc}A\Be&B\Al\\B\De&D\Be\end{array}$&\ig{ohoriz.1}&\ig{odiag.1}&\ig{overt.1}&\ig{tvert.1}&\ig{tdiag.1}&\ig{fhoriz.1}&\ig{ot.1}
\end{tabular}\caption{All double link combinations in the $12$ entry that require analysis}\label{alldoubles}
\end{table}
\end{center}

To exemplify the method of construction of Lax pairs with double links in the $12$ entry, we choose link combination in table \ref{alldoubles} where there is a double link between $A\Be$, $B\De$ and $B\Al$ and a single link between $B\De$ and $D\Be$. Spectral terms that generate this choice of links are given in equation \eqn{ohthapspec12}.
\begin{equation}\label{ohthapspec12}
\begin{array}{lcl}
A=F_1/\Be,&\quad&\Al=F_1/B\\
\De=(F_1+kF_2)/B,&\quad&D=F_2/\Be
\end{array}
\end{equation}
where our usual nomenclature applies, \emph{i.e.} $F_i=F_i(n)$, $k$ = constant.

Using the expressions found in the $12$ entry, the terms that arise in the $21$ entry are $F_1 C/B$, $F_2\Ga/\Be$, $F_1 \Ga/\Be$ and $F_2 C/B$, where $F_1C/B$ arises twice. It is convenient to rearrange the terms found in the $21$ entry into columns of terms that cannot be linked, this is done in equation \eqn{doublerearrange}.
\begin{equation}\label{doublerearrange}
\begin{array}{cc}
F_1 C/B&F_1 \Ga/\Be\\F_2 \Ga/\Be&(F_1+F_2)C/B
\end{array}\quad\rightsquigarrow\quad
\begin{array}{cc}
F_1 \Ga/\Be&F_1 C/B\ig{double.1}\\
F_2 \Ga/\Be&F_2C/B\end{array}
\end{equation}
$F_2\Ga/\Be$ must link with $F_2C/B$ because linking with $F_1C/B$ would lead to the contradictory $F_1\propto F_2$, since that would also necessitate a link between $F_2C/B$ and $F_1\Ga/\Be$. That leaves two possibilities: $B\Ga\propto C\Be$, or we can split $F_2C/B \propto (F_1+F_2)\Ga/\Be$, noting that the multiplicity of the term $F_1C/B$ means that it need not link to another in this entry. The second possibility is neglected, though, as it gives rise to a zero lattice term in one of the diagonal entries to be considered below. Thus, by a process of elimination, the links selected in the $12$ entry leave only one choice for the $21$ entry, which is shown in equation \eqn{double2double}
\begin{equation}\label{double2double}
\begin{array}{ccccccccccc}
\fbox{12}&A\Be&            &B\Al&\quad&&\quad&\fbox{21}&C\Al&            &A\Ga\\
         &    &\ig{ohoriz.1}&    &&\Rightarrow&&        &    &\ig{thoriz.1}&    \\
         &B\De&            &D\Be&&           &&        &D\Ga&            &C\De
\end{array}
\end{equation}

Move now to the $11$ entry where the relevant spectral term products are $A\Al\propto F_1^2/(B\Be)$, and $B\Ga\propto C\Be$, see equation \eqn{otapgroups11}. We choose to link these two spectral terms to get one equation in the $11$ entry and find that there are necessarily two equations in the $22$ entry. Note that the opposite case where there are two equations in the $11$ entry and one in the $22$ entry can also exist by splitting up $B\Ga$ into two terms, however this will lead to the same evolution equation.
\begin{equation}
\begin{array}{ccccccccc}
\fbox{11}&&&& & &| && \\
&&(\as\al-a\ala)&F_1^2/(B\Be)& & &| &(c\bea-\bs\ga)&B\Ga \\
&&&& & &| && \\
 &&&& & & && \\
\fbox{22}&&&F_1F_2/(B\Be)& & &| && \\
&&(\ds\de-d\dea)&& & &| &(b\gaa-\cs\be)&B\Ga \\
&&&F_2^2/(B\Be)& & &| &&
\end{array}\label{otapgroups11}
\end{equation}

Using our symbolism, the links that define this Lax pair are shown in table \ref{otaplinks}.
\begin{table}[h!]
\renewcommand{\arraystretch}{2}
\begin{center}
\begin{tabular}{|c| c |c |c|}
\hline
$\begin{array}{c} \fbox{12}\\
\begin{array}{cc}\mbox{\footnotesize{$A\Be$}}&\mbox{\footnotesize{$B\Al$}}\\
\mbox{\footnotesize{$B\De$}}&\mbox{\footnotesize{$D\Be$}}\end{array}\end{array}$ &$\begin{array}{c} \fbox{21}\\
\begin{array}{cc}\mbox{\footnotesize{$C\Al$}}&\mbox{\footnotesize{$D\Ga$}}\\
\mbox{\footnotesize{$A\Ga$}}&\mbox{\footnotesize{$C\De$}}\end{array}\end{array}$&$\begin{array}{c} \fbox{11}\\
\mbox{\footnotesize{$A\Al$}}\begin{array}{c}\mbox{\footnotesize{$B\Ga$}}\\
\mbox{\footnotesize{$C\Be$}}\end{array}\end{array}$ &$\begin{array}{c} \fbox{22}\\
\mbox{\footnotesize{$D\De$}}\begin{array}{c}\mbox{\footnotesize{$B\Ga$}}\\
\mbox{\footnotesize{$C\Be$}}\end{array}\end{array}$\\
\hline\ig{ohoriz.1}&\ig{thoriz.1}&\ig{dall.1}&\ig{ssplit.1}\\
\hline
\end{tabular}\caption{Links for a Lax pair with double links in the off-diagonal entries}\label{otaplinks}
\end{center}
\end{table}

The spectral term relations that must be satisfied to bring about the links in table \ref{otaplinks} are given in \eqn{otapspecs}.
\begin{equation}\label{otapspecs}
\begin{array}{ccccccc}
A&=&F_1,&\quad\quad&\Al&=&F_1\\
B&=&1,&\quad\quad&\Be&=&1\\
C&=&F_1^2,&\quad\quad&\Ga&=&F_1^2\\
D&=&F_2,&\quad\quad&\De&=&F_1+kF_2\\
\end{array}
\end{equation}
where $F_1$ and $F_2$ are any functions of the spectral variable $n$, such that $F_1\neq kF_2$, $k=$ constant. Notice that a constant appears in the expression for $\De$ in equation \eqn{otapspecs}, since one of the constants of proportionality in this expression cannot be absorbed into the multiplying lattice term, $\de$.

The lattice term equations that arise via the compatibility condition from this Lax pair are shown in equation \eqn{otapeqns}.
\begin{equation}\label{otapeqns}
\begin{array}{rcl}
\as\al-a\ala&=&c\bea-\bs\ga\\
\ds\de-d\dea&=&0\\
b\gaa-\cs\be&=&0\\
\as\be+\bs\de&=&b\ala\\
\bs\de&=&kd\bea\\
\cs\al&=&d\gaa+c\dea\\
\as\ga&=&kc\dea
\end{array}
\end{equation}

We thus conclude the description of the formation of Lax pairs with at least one double link in the $12$ entry. Naturally there are other Lax pairs of this type but they are formed in a similar manner to that described here.

\subsubsection{A triple link in the $12$ entry}
Lastly, link combinations including triple, and possibly double and single links in the $12$ entry must be considered. It is not difficult to see that the only possibility that needs to be investigated is that with one triple link between all four lattice terms in the $12$ entry, any additional links in this entry constrain the problem too heavily and lead to trivial evolution equations only.

A triple link in the $12$ entry of the compatibility condition can be formed by setting the spectral terms to those in equation \eqn{a12} below.
\begin{equation}\label{a12}
\begin{array}{lcl}
A=F_1/\Be,&\quad&\Al=F_1/B\\
\De=F_1/B,&\quad&D=F_1/\Be
\end{array}\rightsquigarrow
\begin{array}{ccccc}
\fbox{12}&A\Be&&B\Al\\
&&\ig{all.1}&\\
&B\De&&D\Be
\end{array}
\end{equation}

Given the values in \eqn{a12}, the resulting spectral terms that appear in the $21$ entry are as shown in equation \eqn{a21}.
\begin{equation}\label{a21}
\begin{array}{ccc}
\fbox{21}&C\Al&A\Ga\\
&D\Ga&C\De\end{array}\rightsquigarrow
\begin{array}{cc}
F_1C/B&F_1\Ga/\Be\\
F_1\Ga/\Be&F_1C/B
\end{array}
\end{equation}
We therefore have two possibilities in the $21$ entry depending on weather or not $B\Ga\propto C\Be$. If $B\Ga$ is proportional $C\Be$ then there is another triple link in the $21$ entry, otherwise there is a pair of single links. The latter case yields the links given in item 4 of table \ref{listofpossibles} and leads to a trivial evolution equation, while the former yields Lax pairs that include one for the LMKdV$_2$ system.

We continue the analysis here assuming $B\Ga\propto C\Be$, in this case the spectral terms in the diagonal entries are shown in  equation \eqn{a11}.
\begin{equation}
\begin{array}{ccccccccc}
\fbox{11}&&&& & &| && \\
&&(\as\al-a\ala)&F_1^2/(B\Be)& & &| &(c\bea-\bs\ga)&B\Ga \\
&&&& & &| && \\
 &&&& & & && \\
\fbox{22}&&&& & &| && \\
&&(\ds\de-d\dea)&F_1^2/(B\Be)& & &| &(b\gaa-\cs\be)&B\Ga \\
&&&& & &| &&
\end{array}\label{a11}
\end{equation}
Again, two possibilities present themselves, this time depending on whether $F_1^2/(B\Be)\propto B\Ga$. The case where the proportionality does not hold provides a Lax pair for the LMKdV$_2$ system, which is discussed in section \ref{LMSG}. When $F_1^2/(B\Be)\propto B\Ga$ does hold, a unique situation unfolds where each entry of the compatibility condition contains only one equation. This special case is discussed in section \ref{aaaa}.

\subsection{List of link combinations}
We are now in a position to tabulate the results. Table \ref{listofpossibles} contains a representative selection of all possible link combinations for $2\times2$ Lax pairs with a single, separable term in each entry of the $L$ and $M$ matrices. There are still other link combinations that were analyzed but do not appear in table \ref{listofpossibles} because they are equivalent to a combination that does appear, or because it is clear that the corresponding Lax pair cannot yield an interesting evolution equation since a similar, less constrained combination of links is listed as trivial or over-determined.
\begin{center}
\begin{table}[!h]
\renewcommand{\arraystretch}{1.75}
\begin{tabular}{|c|c c c c|c|}
\hline&$\begin{array}{c} \fbox{12}\\
\begin{array}{cc}\mbox{\footnotesize{$A\Be$}}&\mbox{\footnotesize{$B\Al$}}\\
\mbox{\footnotesize{$B\De$}}&\mbox{\footnotesize{$D\Be$}}\end{array}\end{array}$ &$\begin{array}{c} \fbox{21}\\
\begin{array}{cc}\mbox{\footnotesize{$C\Al$}}&\mbox{\footnotesize{$D\Ga$}}\\
\mbox{\footnotesize{$A\Ga$}}&\mbox{\footnotesize{$C\De$}}\end{array}\end{array}$&$\begin{array}{c} \fbox{11}\\
\mbox{\footnotesize{$A\Al$}}\begin{array}{c}\mbox{\footnotesize{$B\Ga$}}\\
\mbox{\footnotesize{$C\Be$}}\end{array}\end{array}$ &$\begin{array}{c} \fbox{22}\\
\mbox{\footnotesize{$D\De$}}\begin{array}{c}\mbox{\footnotesize{$B\Ga$}}\\
\mbox{\footnotesize{$C\Be$}}\end{array}\end{array}$&Evolution Eqn\\
\hline
1& \ig{cross.1}&\ig{cross.1}&\ig{dall.1}&\ig{dall.1}&LSG$_2$ \eqn{resLSG}\\
2&\ig{all.1}&\ig{all.1}&\ig{pairs.1}&\ig{pairs.1}&LMKdV$_2$\eqn{resLMKdV}\\
3&\ig{cross.1}&\ig{cross.1}&\ig{pairs.1}&\ig{pairs.1}&Trivial\\
4& \ig{cross.1}&\ig{all.1}&\ig{split.1}&\ig{split.1}&Trivial\\
5&\ig{cross.1}&\ig{cross.1}&\ig{split.1}&\ig{split.1}&Zero term\\
6&\ig{horiz.1}&\ig{horiz.1}&\ig{dall.1}&\ig{pairs.1}&Trivial\\
7& \ig{vert.1}&\ig{vert.1}&\ig{dall.1}&\ig{dall.1}&Underdetermined\\
8&\ig{overt.1}&\ig{tvert.1}&\ig{sall.1}&\ig{dall.1}&Trivial\\
9& \ig{ohoriz.1}&\ig{thoriz.1}&\ig{dall.1}&\ig{sall.1}&Trivial\\
10& \ig{odiag.1}&\ig{tdiag.1}&\ig{sall.1}&\ig{dall.1}&Overdetermined\\
11&\ig{tvert.1}&\ig{overt.1}&\ig{sall.1}&\ig{dall.1}&Trivial\\
12&\ig{tdiag.1}&\ig{odiag.1}&\ig{sall.1}&\ig{dall.1}&Overdetermined\\
13&\ig{fhoriz.1}&\ig{thhoriz.1}&\ig{ssplit.1}&\ig{dall.1}&Overdetermined\\
14&\ig{ot.1}&\ig{ot.1}&\ig{tall.1}&\ig{dall.1}&Overdetermined\\
15&\ig{all.1}&\ig{all.1}&\ig{dall.1}&\ig{dall.1}&Special case\\
\hline
\end{tabular}
\caption{List of link combinations used to construct possible Lax pairs}\label{listofpossibles}
\end{table}
\end{center}
Items 1 and 2 in table \ref{listofpossibles} are analyzed thoroughly in section \ref{LMSG}. The other representative link combinations are dealt with in section \ref{badevos}.

\section{Derivation of higher order LSG and LMKdV systems}\label{LMSG}
While fourteen potentially viable types of Lax pairs were identified in section \ref{links}, only two types lead to non-trivial, well determined evolution equations. The two systems thus found are LMKdV$_2$ and LSG$_2$ and this section describes the derivation of these systems from the general form of their Lax pairs. It is important to note that no freedom in the lattice terms is lost through this process, all values that the lattice terms take are dictated by the sets of equations effectuated by the compatibility condition. As such we conclude that the systems so derived (or equivalent systems) are the most general ones that can be associated with their Lax pairs.

In fact, such calculations have been performed many times before and yet neither LSG$_2$ or LMKdV$_2$ appear to have been published, despite coming from Lax pairs with simple forms that have certainly already been considered previously \cite{srh,sc03}. Hence, it is necessary to outline the method used to derive LSG$_2$ and LMKdV$_2$ in detail.

\subsection{LSG$_2$} \label{LSG2derivation}
For LSG$_2$ the Lax pair used as a starting point is
\begin{subequations}\label{genLSG2LP}
\begin{eqnarray}
L&=&\left(\begin{array}{cc}F_1 a&F_2 b\\F_2 c&F_1 d\end{array}\right)\\
M&=&\left(\begin{array}{cc}F_2 \al&F_1 \be\\F_1 \ga&F_2 \de\end{array}\right)
\end{eqnarray}
\end{subequations}
Where $a$, $b$, $c$, $d$, $\al$, $\be$, $\ga$ and $\de$ are all functions of both the lattice variables $l$ and $m$, referred to as lattice terms, and $F_1$ and $F_2$ depend on the spectral variable only, we refer to these terms as spectral terms.

It is possible to remove some of the lattice terms using a gauge transformation. While this would reduce the complexity of the system, it is not clear at this point which of the lattice terms would best be removed. Experience shows that natural transformations present themselves in the course of solving the compatibility condition and so we shall wait until later to remove some lattice terms, keeping in mind that we expect some freedom to disappear from each of the Lax matrices $L$ and $M$.

The compatibility condition for \eqn{genLSG2LP}, $\widehat{L}M=\ol{M}L$, leads to the following system of difference equations in the lattice terms
\begin{subequations}\label{ccaaCC}
\begin{eqnarray}
\as\al+\bs\ga&=&a\ala+c\bea\label{ccaa1111}\\
\ds\de+\cs\be&=&d\dea+b\gaa\label{ccaa1221}\\
\as\be&=&d\bea\label{ccaa1121}\\
\bs\de&=&b\ala\label{ccaa1122}\\
\cs\al&=&c\dea\label{ccaa1211}\\
\ds\ga&=&a\gaa\label{ccaa1212}
\end{eqnarray}
\end{subequations}

Some of equations \eqn{ccaaCC} are linear and some nonlinear. The linear equations can be solved easily when in the form
\begin{equation}\label{genint}
k_1\hat{\phi}-k_2 \phi=k_3 \bar{\psi}-k_4 \psi
\end{equation}
where $\phi$ and $\psi$ are lattice terms and $k_i$ constants. Equation \eqn{genint} implies
\begin{eqnarray}
\psi&=&k_1\hat{v}-k_2 v+\mu(\frac{k_4}{k_3})^l\nonumber\\
\phi&=&k_3\bar{v}-k_4 v+\la(\frac{k_2}{k_1})^m\nonumber
\end{eqnarray}
where we have introduced the new lattice term $v=v(l,m)$, and $\la=\la(l)$ and $\mu=\mu(m)$ are constants of integration. The same fact also applies in a multiplicative sense, in particular
\begin{equation}
\frac{\widehat{\phi}}{\phi}=\frac{\bar{\psi}}{\psi}\quad\Rightarrow\quad \phi=\la\frac{\bar{v}}{v}, \,\, \psi=\mu\frac{\hat{v}}{v}\nonumber
\end{equation}

We now proceed to solve the system \eqn{ccaaCC}. Equations \eqn{ccaa1121} to \eqn{ccaa1212} are linear and may solved in pairs as follows. Multiply equations
\eqn{ccaa1121} by \eqn{ccaa1212} to find
$\as\ds/(ad)=\bea\gaa/(\be\ga)$, which is integrated for
\begin{eqnarray}
a&=&\la_0^2\bar{v}_5/(v_5 d)\nonumber\\
\be&=&\mu_5\hat{v}_5/(v_5 \ga)\nonumber
\end{eqnarray}
where we have introduced $\la_0=\la_0(l)$, $\mu_5=\mu_5(m)$ and
$v_5=v_5(l,m)$. Now use equation \eqn{ccaa1121} again to find $\bar{v}_5\gaa/(v_5\ga)=\frac{\ds}{\hat{\la}_0}\frac{d}{\la_0}$ and integrate for
\begin{eqnarray}
d&=&\la_0\rho\bar{v}_2/v_2\nonumber\\
\ga&=&\mu_2\hat{v}_2 v_2/v_5\nonumber
\end{eqnarray}
where $\rho=\la_3(l)^{(-1)^m}$. Substitute these values back into the expressions for $a$ and $\be$, and replace $v_5/v_2\mapsto\vo$, and $\mu_5/\mu_2\mapsto\mu_1$ resulting in
\begin{subequations}\label{ccaa1v1}
\begin{eqnarray}
a&=&\frac{\la_0}{\rho}\frac{\voa}{\vo}\\
d&=&\la_0\rho\frac{\vta}{\vt}\\
\be&=&\mu_1\frac{\vos}{\vt}\\
\ga&=&\mu_2\frac{\vts}{\vo}
\end{eqnarray}
\end{subequations}

Perform similar calculations on equations \eqn{ccaa1122} and
\eqn{ccaa1211} to find
\begin{subequations}\label{ccaa1v2}
\begin{eqnarray}
b&=&\la_1\frac{\vtha}{\vf}\\
c&=&\la_2\frac{\vfa}{\vth}\\
\al&=&\frac{\mu_0}{\sigma}\frac{\vths}{\vth}\\
\de&=&\mu_0\sigma\frac{\vfs}{\vf}
\end{eqnarray}
\end{subequations}
Where $\sigma=\mu_3(m)^{(-1)^l}$.

When equations \eqn{ccaa1v1} and \eqn{ccaa1v2} are substituted
into \eqn{ccaa1111} and \eqn{ccaa1221} we find the following equations respectively
\begin{subequations}
\begin{eqnarray}
\la_0\rho\frac{\mu_0}{\sigma}\frac{\voas\vths}{\vos\vth}+\la_1\mu_1\frac{\vts\vths}{\vo\vfs}&=& \frac{\la_0}{\rho}\mu_0\sigma\frac{\voa\vthas}{\vo\vtha}+\la_2\mu_2\frac{\voas\vfa}{\vta\vth}\quad\quad\label{ccaa1112}\\
\frac{\la_0}{\rho}\mu_0\sigma\frac{\vtas\vfs}{\vts\vf}+\la_2\mu_2\frac{\vfas\vos}{\vths\vt}&=& \la_0\rho\frac{\mu_0}{\sigma}\frac{\vts\vfas}{\vt\vfa}+\la_1\mu_1\frac{\vtas\vtha}{\voa\vf}\quad\quad\label{ccaa1222}
\end{eqnarray}
\end{subequations}
Multiplying \eqn{ccaa1112} by $\vo/\voas$ and \eqn{ccaa1222} by $\vf/\vfas$ indicates that certain variables always appear in combination. As such, we set $\vo,\vt\equiv1$ without loss of generality, and rename $\vth= x$ and $\vf= y$. This is the manifest reduction in freedom that was expected from the perspective of gauge transformations. The parameter functions similarly appear in ratios, hence we set $\la_0\equiv1$ and $\mu_0\equiv1$ without loss of generality. Making the substitutions we arrive at a pair of nonlinear partial difference equations in $x$ and $y$, with arbitrary non-autonomous terms $\la_i(l)$ and $\mu_i(m)$, which together form LSG$_2$, the evolution equation associated with the Lax pair.
\begin{subequations}
\begin{eqnarray}
\frac{\rho}{\sigma}\frac{\xs}{x}+\la_1 \mu_1\xas\ys&=&\frac{\sigma}{\rho}\frac{\xas}{\xa}+\frac{\la_2\mu_2}{x\ya} \\
\frac{\sigma}{\rho}\frac{\yas}{\ys}+\frac{\la_2\mu_2}{\xs y}&=& \frac{\rho}{\sigma}\frac{\ya}{y}+\la_1\mu_1\xa\yas
\end{eqnarray}
\end{subequations}
This pair of equations can be thought of as a higher order lattice sine-Gordon system because the lattice sine-Gordon equation (LSG) is
retrieved by setting $y=x$ in either expression
\[
{\rm LSG:}\quad \xas x(\frac{\sigma}{\rho}-\la_1\mu_1\xa\xs)=\frac{\rho}{\sigma}\xa\xs-\la_2 \mu_2
\]
The Lax pair \eqn{LSG2LP} for LSG$_2$ is obtained by substituting the calculated values of the lattice terms back into \eqn{genLSG2LP}.

\subsection{LMKdV$_2$}\label{lmkdv2derivation}
The the general form of the Lax pair for LMKdV$_2$ is similar to that for LSG$_2$, the only difference being that here both matrices exhibit the same dependence on the spectral variable, while the dependence was antisymmetric in the previous case.
\begin{eqnarray}\label{genLMKdV2LP}
L&=&\left(\begin{array}{cc}F_1 a&F_2 b\\F_2 c&F_1 d\end{array}\right)\\
M&=&\left(\begin{array}{cc}F_1 \al&F_2 \be\\F_2 \ga&F_1 \de\end{array}\right)
\end{eqnarray}
As above, $a$, $b$, $c$, $d$, $\al$, $\be$, $\ga$ and $\de$ are all functions of the both the lattice variables $l$ and $m$, $F_1$ and $F_2$ depend on the spectral variable only.

The compatibility condition leads to six equations coming from the different orders of the spectral variable in each entry.
\begin{subequations}\label{aaccCC}
\begin{eqnarray}
\as\al&=&a\ala\label{aacc1111}\\
\bs\ga&=&c\bea\label{aacc1112}\\
\ds\de&=&d\dea\label{aacc1221}\\
\cs\be&=&b\gaa\label{aacc1222}\\
\as\be+\bs\de&=&b\ala+d\bea\label{aacc1121}\\
\cs\al+\ds\ga&=&a\gaa+c\dea\label{aacc1211}
\end{eqnarray}
\end{subequations}
Equations \eqn{aacc1111} and \eqn{aacc1221} are integrated immediately, while \eqn{aacc1112} and \eqn{aacc1222} are multiplied together and dealt with by a similar method to that used for the LSG$_2$ system of section \ref{LSG2derivation} leading to the following results
\begin{subequations}\label{LMKdVv}
\begin{eqnarray}
a&=&\la_1 \voa/\vo\\
b&=&\la_0\rho\vtha/\vf\\
c&=&\la_0\vfa/(\rho\vth)\\
d&=&\la_2\vta/\vt\\
\al&=&\mu_1\vos/\vo\\
\be&=&\mu_0\vths/(\sigma\vf)\\
\ga&=&\mu_0\sigma\vfs/\vth\\
\de&=&\mu_2\vts/\vt
\end{eqnarray}
\end{subequations}
Where $v_i=v_i(l,m)$, $\la_i=\la_i(l)$, $\mu_i=\mu_i(m)$, $\rho=\la_3^{(-1)^m}$ and $\sigma=\mu_3^{(-1)^l}$. Substituting these values into the two remaining equations, \eqn{aacc1121} and \eqn{aacc1211}, shows that terms consistently appear in ratios, as they did with the LSG$_2$ system. We choose to set $\vt\equiv\vth\equiv1$, $\la_0\equiv\mu_0\equiv1$, $\vo=x$ and $\vf=y$ and arrive at LMKdV$_2$:
\begin{eqnarray}\label{LMKdV2evo}
\frac{\la_1}{\sigma}\frac{\xas}{\xs}+\frac{\mu_2}{\rho}\frac{y}{\ys}&=& \la_2\sigma\frac{y}{\ya}+\rho\mu_1\frac{\xas}{\xa}\label{LMKdVevo1}\\
\rho\mu_1\xs\yas+\la_2\sigma x\ys&=& \frac{\mu_2}{\rho}x\ya+\frac{\la_1}{\sigma}\xa\ys\label{LMKdVevo2}
\end{eqnarray}

\section{How most link combinations lead to bad evolution equations}\label{badevos}
Here we explain how most of the potentially viable Lax pairs fail to produce interesting evolution equations. The topic is split into four parts dealing with Lax pairs that lead to trivial, over-determined, under-determined evolution equations and the special case of item 15 in table \ref{listofpossibles}.

\subsection{Lax pairs that yield only trivial evolution equations}\label{trivial}
This section is pertinent to items 3, 4, 6, 8, 9 and 11 in table \ref{listofpossibles}, those Lax pairs who's compatibility conditions can be solved to a point where only linear equations remain, or the equations can be reduced to first order equation in one lattice direction only.

The simplest route to triviality is to have a Lax pair with a set of equations that are all linear, as per item 3 in table \ref{listofpossibles}. The compatibility condition leads to a set of eight linear equations in the eight initial lattice terms.
\begin{equation}\label{ccppCC}
\begin{array}{ccccccc}
\as\al&=&a\ala,&\quad\quad&\bs\ga&=&c\bea\\
\ds\de&=&d\dea,&\quad\quad&\cs\be&=&n\gaa\\
\as\be&=&d\bea,&\quad\quad&\bs\de&=&b\ala\\
\ds\ga&=&a\gaa,&\quad\quad&\cs\al&=&c\dea
\end{array}
\end{equation}

Equations \eqn{ccppCC} can be solved easily using techniques described in section \ref{LMSG}, with the result of a simple, linear evolution equation. However, it is not necessary to conduct such analysis on this system since all equations \eqn{ccppCC} are linear and they can not be expected to produce a nonlinear evolution equation. For this reason, other examples of link combinations that produce only linear equations have been omitted from table \ref{listofpossibles}.

Item number 8 from table \ref{listofpossibles} is an example of a Lax pair that leads to a trivial evolution equation in a more complex way. The equations that come out of its compatibility condition are
\begin{subequations}
\begin{eqnarray}
\as\al&=&a\ala\label{ovtvsa111}\\
\bs\ga&=&c\bea\label{ovtvsa112}\\
\ds\de+\cs\be&=&d\dea+b\gaa\label{ovtvsa221}\\
\as\be+\bs\de&=&b\ala\label{ovtvsa121}\\
kb\ala&=&-d\bea\label{ovtvsa122}\\
\cs\al&=&a\gaa+c\dea\label{ovtvsa211}\\
k\cs\al&=&-\ds\ga\label{ovtvsa212}
\end{eqnarray}
\end{subequations}
Where $k$ is a constant. This system of equations is solved as follows: integrate \eqn{ovtvsa111} for $a=\la\va/v$, $\al=\mu\vs/v$, introducing $\la=\la(l)$, $\mu=\mu(m)$ and $v=v(l,m)$. Multiply equations \eqn{ovtvsa112} and \eqn{ovtvsa212}, then divide the product by \eqn{ovtvsa122} and use the values calculated for $a$ and $\al$ to find
\begin{equation}\label{ovtvsa1111}
\frac{\bs\cs\vs}{bcv}=\frac{\ds\vas}{d\va}
\end{equation}
Equation \eqn{ovtvsa1111} is integrated for $b$, which is substituted into equations \eqn{ovtvsa122} and \eqn{ovtvsa112} yielding the following values
\begin{eqnarray}
b&=&\la_2\frac{d\va}{cv}\nonumber\\
\bea&=&-k\mu\la_2\frac{\vas}{cv}\nonumber\\
\ga&=&-k\mu\frac{\cs\vs}{\ds v}\nonumber
\end{eqnarray}
There are now three equations left to solve, \eqn{ovtvsa221}, \eqn{ovtvsa121} and \eqn{ovtvsa211}, where it is found that $c$ and $v$ always appear as a product which suggests we introduce $u=cv$. Equation \eqn{ovtvsa221} is then used to write
\begin{equation}\nonumber
\dea=\mu\frac{\hat{u}}{u}+k\mu\la\frac{\hat{\bar{u}}}{u\das}
\end{equation}
Finally, we introduce $x=\ul{u}d/u$ and achieve the final two equations in $x$
\begin{subequations}\label{ovtvsaevo}
\begin{eqnarray}
\la_2(\xs-x)&=&k(\ul{\la}_2\la-\ul{\la}\la_2)\label{ovtvsaevo1}\\
\xs-x+k(\ul{\la}-\la_2)&=&k\frac{x}{\xas}(\mu\la-\la_2)\label{ovtvsaevo2}
\end{eqnarray}
\end{subequations}
Equations \eqn{ovtvsaevo} is an overdetermined set of two equations in the one variable, $x$, which cannot hope to yield an interesting evolution equation. This is because \eqn{ovtvsaevo1} can be used to remove the dependence of $x$ on $m$ ($m$ is the independent variable of the `\,$\hat{ \, \,}$\,' direction), leaving $x$ with a first order dependence on $l$ in \eqn{ovtvsaevo2} at best.

\subsection{Overdetermined Lax pairs} \label{od}
Here items 10 and 12 through 14 in table \ref{listofpossibles} are dealt with. Such Lax pairs have compatibility conditions that boil down to more equations than there are free lattice terms. These are not necessarily trivial, as the solution to one equation may solve another as well, that is it may be possible to make one or more equations redundant. Or one equation may be a compatible similarity condition for another equation as explained in \cite{nc95}. Some systems of this type, where any hope of supporting an interesting evolution equation has been quashed, have already been considered in section \ref{trivial}. The remaining over-determined systems are considered here, however we do not attempt to resolve the issue of whether these systems support interesting evolution equations, we simply list them as being overdetermined.

The most interesting instances of this type of system arise from Lax pairs with two double links in the off-diagonal entries, those represented by item 14 from table \ref{listofpossibles}. All such systems, with minimum constraint on the lattice terms in the diagonal entries of the compatibility condition, lead to the same evolution equations. An example of a Lax pair with two double links in the off-diagonal entries is
\begin{subequations}\label{ddpaLP}
\begin{eqnarray}
L&=&\left(\begin{array}{cc}F_1a&b\\(F_1^2+k_4F_1F_2)c&F_2d\end{array}\right)\\
M&=&\left(\begin{array}{cc}(F_1+k_2F_2)\al&\be\\(F_1^2+k_3F_1F_2)\ga&(F_1+k_1F2)\de\end{array}\right)
\end{eqnarray}
\end{subequations}
where lower case letters except $k_i$ are lattice terms, $k_i$ are constants of proportionality and $F_i$ are spectral terms with $F_1$ not proportional to $F_2$. This particular Lax pair is especially interesting because the solution of its compatibility condition involves integrating both additive and multiplicative linear difference equations, as described near equation \eqn{genint} in section \ref{LMSG}. The final evolution equations achieved are relatively complicated and nonlinear, however there are two equations for one variable and it remains to be seen whether they can be reconciled. The following outlines how to solve the compatibility condition.

In the compatibility condition, $\widehat{L}M=\ol{M}L$, the $11$ entry dictates $k_2=k_3=k_4$. Redefine $F_2\mapsto k_2F_2$ to remove $k_2$ from everywhere except the $22$ entry of $L$ where we set $k_0=1/K_2$ and thus arrive at the following equations from the various orders of the compatibility condition
\begin{subequations}
\begin{eqnarray}
\as\al+\bs\ga&=&a\ala+c\bea\label{ddpa111}\\
\cs\bea&=&b\gaa\label{ddpa221}\\
\ds\de&=&d\dea\label{ddpa222}\\
\as\be+\bs\de&=&b\ala\label{ddpa121}\\
k_1\bs\de&=&b\ala+k_0d\bea\label{ddpa122}\\
\cs\al&=&a\gaa+c\dea\label{ddpa211}\\
\cs\al+k_0\ds\ga&=&k_1c\dea\label{ddpa212}
\end{eqnarray}
\end{subequations}
Integrate \eqn{ddpa222} for $d=\la\va/v$, $\de=\mu\vs/v$, introducing $v=v(l,m)$, $\la=\la(l)$ and $\mu=\mu(m)$, and use equation \eqn{ddpa221} to see that $\be=b\gaa/\cs$. Now define $s=c/\va$, $t=\ga/\vs$ and $u=bv$ so that
\begin{subequations}
\begin{eqnarray}\label{ddpaa}
&&\be=b \bar{t}/\hat{s}\\
\eqn{ddpa212}&\Rightarrow&\al=\frac{1}{\hat{s}}(k_1\mu s-k_0\la t)\\
\eqn{ddpa211}&\Rightarrow&a=\frac{1}{\bar{t}}((k_1-1)\mu s-k_0\la t)
\end{eqnarray}
\end{subequations}
Substituting these \eqn{ddpaa} values into equation \eqn{ddpa122} and rearranging leads to
\begin{equation}
k_1\widehat{\left(\frac{u\bar{s}}{\la\laa}\right)}-k_1\left(\frac{u\bar{s}}{\la\laa}\right)= k_0\ol{\left(\frac{u\bar{t}}{\la\mu}\right)}-k_0\left(\frac{u\bar{t}}{\la\mu}\right)
\end{equation}
which is an additive linear difference equation that can be integrated to find
\begin{subequations}\label{sumresult}
\begin{eqnarray}
\frac{u\bar{s}}{\la\laa}&=&\bar{w}-w+\la_2\\
\frac{u\bar{t}}{\la\mu}&=&\hat{w}-w+\mu_2
\end{eqnarray}
\end{subequations}

The two equations that remain, equations \eqn{ddpa111} and \eqn{ddpa121} respectively, are written in terms of $s$, $t$ and $u$
\begin{subequations}\label{ddpaevo}
\begin{eqnarray}
\frac{\hat{\bar{s}}}{\hat{\bar{t}}}((k-1)\hat{\mu}-k_0\la\frac{\hat{t}}{\hat{s}})(k_1\mu\frac{s}{t}-k_0\la)+\hat{u}\hat{\bar{s}}&=& ((k-1)\mu\frac{s}{t}-k_0\la)(k_1\mu-k_0\bar{\la}\frac{\bar{t}}{\bar{s}})+\bar{u}\bar{\bar{t}}\frac{s}{t}\quad\quad\quad\quad\label{ddpaevo1}\\
\frac{\hat{\bar{s}}}{\hat{\bar{t}}}((k-1)\hat{\mu}-k_0\la\frac{\hat{t}}{\hat{s}})+\mu\frac{\hat{u}\hat{\bar{s}}}{u\bar{t}}&=&k_1\mu\frac{\bar{s}}{\bar{t}}-k_0\laa\label{ddpaevo2}
\end{eqnarray}
\end{subequations}
Make a further change of variables $x=u\bar{t}$, $y=u\bar{s}$, to completely remove $u$ from equations \eqn{ddpaevo} and \eqn{sumresult}. Equation \eqn{ddpaevo2} becomes
\begin{equation}
\mu\frac{\hat{\bar{y}}}{\bar{x}}+\frac{\hat{\bar{y}}}{\hat{\bar{x}}}((k_1-1)\hat{\mu}-k_0\la\frac{\hat{x}}{\hat{y}})=k_1\mu\frac{\bar{y}}{\bar{x}}-k_0\laa\label{ddpau}
\end{equation}
while equation \eqn{ddpaevo2} becomes
\begin{equation}
\frac{\hat{\bar{y}}}{\hat{\bar{x}}}((k_1-1)\hat{\mu}-k_0\la\frac{\hat{x}}{\hat{y}})(k_1\mu\frac{y}{x}-k_0\la)+\hat{\bar{y}}= ((k_1-1)\mu\frac{y}{x}-k_0\la)(k_1\mu-k_0\laa\frac{\bar{x}}{\bar{y}})+\bar{\bar{x}}\frac{y}{x}\label{ddpaevo2a}
\end{equation}
where $x$ and $y$ are both written in terms of $w$ according to
\begin{subequations}
\begin{eqnarray}
\xa&=&\frac{\la\mu}{k_0}(\hat{w}-w+\mu_2)\\
\ya&=&\frac{\la\laa}{k_1}(\bar{w}-w-\la_2)
\end{eqnarray}
\end{subequations}
Hence, there are two complicated, nonlinear equations for the one variable $w$, \eqn{ddpau} and \eqn{ddpaevo2a}. These two equations may or may not be reconcilable, one way that they could be reconciled is if one equation was shown to be a compatible similarity constraint for the other \cite{np91,nc95}

\subsection{Underdetermined Lax pairs} \label{ud}
There are Lax pairs who's compatibility condition can be solved completely, while still leaving at least one lattice term free. In these cases there is no genuine evolution equation, although the freedom inherent in the system could to cause it to appear as though there was. In fact, any evolution equation, trivial, integrable or even chaotic, could be falsely represented by such a Lax pair. One such case is item 7 on table \ref{listofpossibles} that has as its compatibility condition
\begin{subequations}
\begin{eqnarray}
\as\al+\bs\ga&=&a\ala+c\bea\label{vvaa111}\\
\ds\de+\cs\be&=&d\dea+b\gaa\label{vvaa221}\\
\as\be&=&-\bs\de\label{vvaa121}\\
b\ala&=&-d\bea\label{vvaa122}\\
\cs\al&=&-\ds\ga\label{vvaa211}\\
a\gaa&=&-c\dea\label{vvaa212}
\end{eqnarray}
\end{subequations}
Here is a roadmap to the solution: multiply equation \eqn{vvaa121} by \eqn{vvaa212} and divide by \eqn{vvaa122} and \eqn{vvaa211} to find an expression that can be integrated for
\begin{eqnarray}
d&=&\la_1\va cb/(v a)\nonumber\\
\ga&=&\mu_1\vs\al\de/(v\be)\nonumber
\end{eqnarray}

Substituting these values for $d$ and $\ga$ into the ratio of equations \eqn{vvaa122} and \eqn{vvaa212} shows that $v=\la_1\mu_1\vas$, which indicates that $v$ must be separable into a product such as $v=\la_2(l)\mu_2(m)$, where $\la_1=\la_2/\laa_2$ and $\mu_1=\mu_2/\hat{\mu}_2$. With the above values included and no further integration required, equation \eqn{vvaa121} is used to find $\de$ while \eqn{vvaa122} offers $c$. In summary:
\begin{subequations}\label{vvaaevo}
\begin{eqnarray}
d&=&-b\ala/\bea\label{vvaa1}\\
\ga&=&-\as\al/\bs\\
\de&=&-\as\be/\bs\\
c&=&-a\ala/\bea
\end{eqnarray}
\end{subequations}

The key feature is that, with \eqn{vvaaevo}, equations \eqn{vvaa111} and \eqn{vvaa221} are automatically satisfied, furnishing us with no more constraints on the remaining lattice terms $a$, $b$, $\al$ and $\be$. Therefore, equations \eqn{vvaaevo} are the evolution equations, but these are not uniquely determined and there is freedom enough to write any equation at all into this set, including equations that are known to be not integrable. The undeniable conclusion is that any Lax pair of this type is false because one would expect that the information to be gleaned about the solution to any evolution equation associated with this Lax pair must be underdetermined like equation \eqn{vvaaevo}. Note further that the level of freedom left in the system is exactly that which can be removed by gauge transformations, so, in essence, this systems compatibility is simply determined by the values of the parameter functions and no evolution equation exists.

\subsection{A special case}\label{aaaa}
Item 15 from table \ref{listofpossibles} is a special case that sees all lattice terms linked in all four entries of the compatibility condition. Thus, the compatibility condition supplies only nonlinear equations, none of which can be explicitly integrated, and so no parameter functions present themselves as constants of integration. This is an unusual situation but it is not necessary to solve the system because of the following arguments regarding the spectral dependence.

To form the links that define this Lax pair we require the following conditions, or an equivalent set, on the spectral terms.
\begin{equation}\label{aaaaspecs}
\begin{array}{lcl}
A=B\Ga/\Al&\quad&\Be=\Al^2/\Ga\\
C=B\Ga^2/\Al^2&&\De=\Al\\
D=B\Ga/\Al&&
\end{array}
\end{equation}
Equation \eqn{aaaalp} shows the general form of the Lax pair that possesses the links in question is
\begin{subequations}\label{aaaalp}
\begin{eqnarray}
L&=&BF_1\left(\begin{array}{cc}a&b/F_1\\cF_1&d\end{array}\right)\\
M&=&\Al\left(\begin{array}{cc}\al&\be/F_1\\\ga F_1&\de\end{array}\right)
\end{eqnarray}
\end{subequations}
where $F_1(n)=\Ga/\Al$. The prefactors in equation \eqn{aaaalp} are obsolete because they cancel in the compatibility condition. This leaves the Lax pair with the same dependence on just one spectral term in both the $L$ and $M$ matrices. As such, a gauge transformation can be used to completely remove the dependence on the spectral variable from the linear problem, implying that equation \eqn{aaaalp} is actually not a Lax pair at all.

\section{conclusion}
All $2\times2$ Lax pairs, with one separable term in the four entries of each matrix, have been considered through the various combinations of terms possible in their compatibility conditions. It has been shown that the only non-trivial evolution equations that can be supported are the higher order generalizations of the LMKdV and LSG equations, with the possible exception of over-determined systems that may yet be consistent, see section \ref{od}. The results of the present work support the postulated connection between Lax pairs and singularity confinement \cite{srh}.

There is an important question in where the present work sits in relation to studies in multidimensional consistency, or consistency around a cube (CAC), that provide a method of searching for integrable partial difference equations where a Lax pair can be derived as a bi-product of the procedure \cite{n02,bs}. The present work adds to multidimensional consistency studies by removing the restriction that the Lax matrices must be symmetric, a consequence of using the same $Q$ equation in all three directions of the cube \cite{h04}, and also leads to non-autonomous equations, where multidimensional consistency studies have only considered autonomous systems. Non-autonomy is vital to reductions of the type used in \cite{grswc} and \cite{hay}, that lead to nonlinear ordinary difference equations and Lax pairs for them. In addition, while it has been shown that (CAC) ensures the existence of a Lax pair, the converse is not necessarily true, which in itself indicates a need for the present completeness study.

Future studies using the techniques explained here should investigate Lax pairs with non-separable terms or with more terms in each matrix entry, and possibly aim for a general algorithm for dealing with an arbitrary number of terms in each entry of the Lax matrices. Other types, such as differential difference Lax pairs, should be considered, as should purely continuous Lax pairs where there is already a vast body of knowledge with which to compare results. Further, this technique is easily adaptable to isomonodromy Lax pairs.

There is some conflict about whether the parameters of the $Q_4$ equation, in the ABS scheme, must lie on elliptic curves or not \cite{abs03,as04,v06}. The Lax pair for $Q_4$, found via multidimensional consistency, is for a version of the equation where the autonomous parameters are restricted to elliptic curves \cite{n02}. An exploration into the links present in that Lax pair might resolve the issue by providing a Lax pair for $Q_4$ with non-autonomous, and possibly free, parameters.


\begin{thebibliography}{1}

\bibitem{nc}
F.W. Nijhoff, and H.W. Capel,
\newblock The discrete Korteweg-de Vries equation,
\newblock {\em Acta Appl. Math.},  {\bf 39} (1-3) (1995) 133--158

\bibitem{pgr93}
V. Papageorgiou, B. Grammaticos, and A. Ramani,
\newblock Integrable lattices and convergence acceleration algorithms,
\newblock {Phys. Lett. A} {\bf 179} (1993) 111--115

\bibitem{bbt03}
O.~Babelon, D.~Bernard and M.~Talon,
\newblock {Introduction to classical integrable systems,}
\newblock {Cambridge University Press,} Cambridge 2003

\bibitem{hay2}
M. Hay,
\newblock {Hierarchies of nonlinear integrable $q$-difference equations from series of Lax pairs,}
\newblock {\em J. Phys. A: Math. Gen.}, {\bf 40} No. 34 (2007), 10457--10471

\bibitem{jc}
C. Creswell and N. Joshi,
\newblock {The discrete Painlevé I hierarchy,}
\newblock {\em London Math. Soc. Lecture Notes Series.}, {\bf 255} (1999), 197

\bibitem{srh}
R.~Sahadevan, O.~G.~Resin, and P.~E.~Hydon,
\newblock{Integrability conditions for nonautonomous quad-graph equations,}
\newblock {\em J. Math. Anal. Appl.}, {\bf 331} No.1 (2007) 712--726

\bibitem{n00}
M.~Noumi,
\newblock {Painlev\'e Equations through Symmetry,}
\newblock {Translations of Mathematical Monographs 223}
\newblock American Mathematical Society 2004

\bibitem{akns74}
M.J. Ablowitz and D.J. Kaup and A.C. Newell and H. Segur,
\newblock{The inverse scattering transform --- {Fourier} analysis for nonlinear problems}
\newblock{\em Stud. Appl. Math.} {\bf 53} (1974) 249--315

\bibitem{al76}
M.~Ablowitz and J.~Ladik",
\newblock{Nonlinear differential-difference equations and {Fourier} analysis,}
\newblock{\em J. Math. Phys.}, {\bf 17} (1976) 1011--1018

\bibitem{jbh92}
N.~Joshi, D.~Burtonclay and R.~Halburd,
\newblock{Nonlinear Nonautonomous Discrete Dynamical Systems from a General Discrete Isomonodromy Problem}
\newblock{\em Lett. Math. Phys.}, {\bf 26} (1992) 123--131

\bibitem{as81}
M.~Ablowitz and H.~Segur,
\newblock {Solitons and the inverse scattering transform,}
\newblock {SIAM Studies in Applied Mathematics,} Philidelphia 1981

\bibitem{sc03}
R.~Sahadevan and H.W.~Capel,
\newblock Complete integrability and singularity confinement of nonautonomous modified Korteweg–de Vries and sine Gordon mappings
\newblock {\em Physica A}, {\bf 330} (2003) 373--390

\bibitem{w98}
R.~S.~Ward,
\newblock {Lax pairs for integrable lattice systems,}
\newblock {\em J. Math. Phys.}, {\bf 40} No. 1 (1999) 299--308

\bibitem{h04}
J.~Hietarinta,
\newblock{A new two-dimensional model that is 'consistent around a cube',}
\newblock{\em J. Phys. A: Math. Gen.} {\bf 37} (2004) L67--L73

\bibitem{grswc}
B.~Grammaticos, A.~Ramani, J.~Satsuma, R.~Willox and A.~S.~Carstea,
\newblock {Reductions of integrable lattices,}
\newblock {\em J. Nonlin. Math. Phys.}, {\bf 12} Supp. 1 (2005) 363--371.

\bibitem{hay}
M. Hay, J.~Hietarinta, N. Joshi and F.W.  Nijhoff,
\newblock {A Lax Pair for a lattice mKdV equation, reductions to q-Painleve equations and associated Lax
pairs,}
\newblock {\em J. Phys. A: Math. Gen.}, {\bf 40} No. 2 (2007),
   F61--F73

\bibitem{abs03}
V.E.~Adler, A.I.~Bobenko and Y.B.~Suris
\newblock {Classification of integrable equations on quad-graphs. The consistency approach,}
\newblock {\em Comm. Math. Phys.}, {\bf 233} No. 3 (2003), 513--543

\bibitem{as04}
V.E.~Adler and Y.B.~Suris,
\newblock {$Q_4$: Integrable master equation related to an elliptic curve,}
\newblock {\em International Mathemtics Research Notice} {\bf 2004} No. 47 (2004), 2523--2533

\bibitem{v06}
C.M.~ Viallet,
\newblock {Algebraic entropy for lattice equations,}
\newblock {\em arXiv:math-ph/0609043v2}

\bibitem{n02}
F.W. Nijhoff,
\newblock Lax pair for the Adler (lattice Krichever-Novikov) system,
\newblock {\em Phys. Lett. A}, {\bf 297}(1-2) (2002) 49--58.

\bibitem{cm00}
R.~Conte and M.~Musette,
\newblock {Towards second order Lax pairs to discrete Painlev\'e equations of first degree,}
\newblock {\em Chaos, Solitons and Fractals}, {\bf 11} (2000), 41--52

\bibitem{np91}
F.W. Nijhoff and V.G. Papageorgiou,
\newblock {Similarity reductions of integrable lattices
and discrete analogues of the Painlev\'e II equation,}
\newblock {\em Phys. Lett. A}, {\bf 153} (1991) 337-344

\bibitem{nc95}
F.W. Nijhoff and H.W. Capel,
\newblock{The discrete Korteweg-de Vries equation,}
\newblock{\em Acta Appl. Math.}, {\bf 39} (1995) 133--158

\bibitem{nw01}
F.W. Nijhoff and A.J. Walker,
\newblock {The discrete and continuous painlev\'{e} VI hierarchy and
the Garnier systems,}
\newblock {\em Glasgow Math. J.}, {\bf 43A} (2001) 109-123


\bibitem{bs}
A.I. Bobenko and Y.B. Suris,
\newblock Integrable systems on quad-graphs,
\newblock {\em IMRN} {\bf 11} (2002)  573--611.

\bibitem{N96}
F.~ Nijhoff (1996),
\newblock {\em Discrete Integrable Geometry and Physics,} Eds A Bobenko and R seiler, (Oxford University Press)

\bibitem{KTGR00}
M.~Kruskal, K.~Tamizhmani, B.~Grammaticos and A.~Ramani,
\newblock Asymmetric discrete Painleve equations,
\newblock {\em Reg. Chaot. Dyn.}, {\bf 5} (2000), 273

\bibitem{s}
H.~Sakai, A {$q$}-analog of the {G}arnier system,
\newblock {\em Funkcial. Ekvac.} {\bf 48} (2005) 273--297.

\end{thebibliography}
\end{document}